# Experimental measurement-based quantum computing beyond the cluster-state model

Wei-Bo Gao[*,1], Xing-Can Yao[*,1], Jian-Ming Cai[2,3], He Lu[1], Ping Xu[1], Tao Yang[1], Chao-Yang Lu[1], Yu-Ao Chen[1], Zeng-Bing Chen[1], and Jian-Wei Pan[1,4]

[1]*Hefei National Laboratory for Physical Sciences at Microscale and Department of Modern Physics, University of Science and Technology of China, Hefei, Anhui 230026, China*

[2]*Institut für Quantenoptik und Quanteninformation, Österreichische Akademie der Wissenschaften, Technikerstraße 21A, A-6020 Innsbruck, Austria*

[3]*Institut für theoretische Physik, Universität Innsbruck, Technikerstraße 25, A-6020 Innsbruck, Austria*

[4]*Physikalisches Institut, Ruprecht-Karls-UniversitÄat Heidelberg, Philosophenweg 12, 69120 Heidelberg, Germany*

[*]*These authors contributed equally to this work.*

The paradigm of measurement-based quantum computation opens new experimental avenues to realize a quantum computer and deepens our understanding of quantum physics. Measurement-based quantum computation starts from a highly entangled universal resource state. For years, clusters states have been the only known universal resources. Surprisingly, a novel framework namely quantum computation in correlation space has opened new routes to implement measurement-based quantum computation based on quantum states possessing entanglement properties different from cluster states. Here we report an experimental demonstration of every building block of such a model. With a four-qubit and a six-qubit state as distinct from cluster states, we have realized a universal set of single-qubit rotations, two-qubit entangling gates and further Deutsch's algorithm. Besides being of fundamental interest, our experiment

**proves in-principle the feasibility of universal measurement-based quantum computation without using cluster states, which represents a new approach towards the realization of a quantum computer.**

Measurement-based quantum computation (MQC), also called one-way quantum computation, has generated enormous interest in the quantum information community since its discovery in 2001[1-3]. In this computational model, two steps are required (i) preparing an algorithm-independent universal resource state and (ii) performing single-qubit measurements with classical feed-forward of their outcomes. The universal resources lie at the heart of measurement-based quantum computation. Cluster states are the sole universal resources that people have been studying both theoretically and experimentally[1-9]. Remarkably, it was found recently by Gross and Eisert that, in the framework of MQC in correlation space many novel entangled states can serve as universal resources for MQC, and singular entanglement properties of cluster states can be relaxed for a universal resource[10-12]. In this model, quantum information is stored and processed in a virtual system (i.e. the correlation space), which is mapped to the Hilbert space of physical particles by means of projection measurements.

The discovery has greatly enriched the universal resources for MQC. These novel resource states offer more flexibility to implement quantum computation in various kinds of physical systems, e.g. optical systems, cold atoms and polar molecules in optical lattices[13], or many-body systems in condensed matter physics[14]. One advantage for optical systems is that certain types of resource states can be efficiently prepared using controlled phase gates, the success probability of which increases rapidly for smaller phase values at certain ranges[15-17]. Combined with the other techniques, e.g. integrated waveguides[18-20], the working principles demonstrated here may provide a new element towards a larger scale photonic quantum computing. For optical lattice and condensed-matter systems, in contrast with cluster states, the sources of MQC in correlation space may arise as the natural ground states of a Hamiltonian with only two-body interaction[10-14, 21-24]. The demonstration of MQC in correlation space will stimulate more interest in tailoring novel universal resource states with the features of different physical systems.

With these novel universal resources, one can also gain new insights into the fundamental problem: what are the essential features of quantum states responsible for the speedup of quantum computers over classical devices[25-28]? In MQC, the computational power is embedded in the universal resource states, entanglement properties of which are thus believed to sustain the speedup of quantum computation. For this reason, entanglement properties of cluster states appear to be paramount and favourable for quantum computation, e.g.: two generic qubits are uncorrelated[9], which makes it possible to logically break down a large scale MQC into small components; and every particle is maximally entangled with the rest of the state in order to guarantee deterministic operations. Nevertheless, in contrast to cluster states, universal resource states with arbitrarily small local entanglement do exist, and a non-vanishing correlation length as a matter of fact is not an obstacle for universal MQC[10-12].

In this article we prepare such a new type of multi-photon entangled states different from cluster states and demonstrate the working principles of MQC in correlation space. First, we generate a four-photon four-qubit state entangled with polarized photons to implement single qubit rotations and the strategy to compensate the measurement randomness, which is a distinct feature of MQC in correlation space. Furthermore, a two-qubit entangling gate and Deutsch's algorithm are implemented based on a four-photon six-qubit state entangled in both the polarization and spatial modes. We show that the generated entangled states have entanglement properties different from cluster states by measuring the two-point correlation functions and single-site entropy in the states, although they still satisfy the same entanglement criteria for universality[25, 28]. Our results show that quantum states different from cluster states can also serve as promising candidates for one-way quantum computation, and fundamentally not all entanglement properties of cluster states are indispensable for quantum computation.

## Theoretical description of the resource states

General universal one-dimensional (1D) computational wires are expressed as matrix product states (MPS)[21, 29, 30]

$$|\Psi\rangle = \sum_{s_i} \langle r|A[|s_n\rangle]\cdots A[|s_1\rangle]|l\rangle |s_1\cdots s_n\rangle \qquad (1)$$

where $A[|s_i\rangle]$ are single-qubit operators, $|l\rangle$ and $|r\rangle$ represent left and right boundary vectors in the so-called correlation space, which is the auxiliary two-dimensional vector space the matrices $A[|s_i\rangle]$ act on. A local projective measurement on site $i$ with the outcome $|\phi_i\rangle$ will induce the action of the operator $A[|\phi_i\rangle] = \sum_{s_i} \langle \phi_i|s_i\rangle A[|s_i\rangle]$ in the correlation space, which can serve as the initialization, single qubit rotation and readout of a logical qubit[10-12]. Quantum 1D wires can be coupled to form a two-dimensional (2D) resource state for the implementation of entangling gates[12].

A four-qubit 1D MPS (see Fig. 1a) encoded on the polarized photons is used to demonstrate single-qubit rotations. Here we take the left and right boundary vectors as $|l\rangle = |+\rangle$ and $|r\rangle = |0\rangle$, where $|+\rangle = \frac{1}{\sqrt{2}}(|0\rangle + |1\rangle)$. Moreover, for the first three sites of the state, the tensor matrices are[28]

$$A[|H\rangle] = \hat{H}\cos\theta \quad \text{and} \quad A[|V\rangle] = \hat{H}\hat{Z}\sin\theta. \qquad (2)$$

Here $|H\rangle$ and $|V\rangle$ denote the horizontal and vertical polarization, which represent the qubits $|0\rangle$ and $|1\rangle$; $\hat{H}$ denotes the Hadamard gate and $\hat{Z}$ refers to the Pauli matrix $\sigma_z$. For the end site, we adopt a simple alternative by changing the tensor matrix to $B[|H\rangle] = \hat{H}$, $B[|V\rangle] = \hat{H}\hat{Z}$. Substituting them into equation (1), the four-qubit state is explicitly written as

$$|\psi_4\rangle = c|H\rangle_1 (c|H\rangle + s|V\rangle)_2 (c|H\rangle|P\rangle + s|V\rangle|M\rangle)_{34} \\ + s|V\rangle_1 (c|H\rangle - s|V\rangle)_2 (c|H\rangle|M\rangle + s|V\rangle|P\rangle)_{34}, \quad (3)$$

where $|P\rangle = \frac{1}{\sqrt{2}}(|H\rangle + |V\rangle)$ and $|M\rangle = \frac{1}{\sqrt{2}}(|H\rangle - |V\rangle)$. For simplicity, we denote $c = \cos\theta$ and $s = \sin\theta$ in the above equation and the following text. For any $\theta$ that $\cos\theta \neq 0$ and $\sin\theta \neq 0$, such a MPS can serve as a quantum wire for single qubit logic gates[28]. By tuning the parameter $\theta$, the entanglement properties can be very different. In our experiment we choose the angle $\theta$ as $\pi/6$, and thus $c = \sqrt{3}/2$; $s = 1/2$. In this state, the processing of logical information is implemented in the part of qubits with the tensor matrix $A$; while the last site serves as readout and has the function of mapping the logical information carried by the correlation system to the physical qubit as $|0\rangle_c \rightarrow |P\rangle_p$ and $|1\rangle_c \rightarrow |M\rangle_p$[28].

To demonstrate the two-qubit entangling gate, a 2D resource state is required. As shown in Fig. 1b, our 2D state is constructed by coupling two 1D MPS (1-2-1' and 3-3') with site 4 via a general scheme to construct universal resources from arbitrary computational wires (see Supplementary Information). The coupling is implemented by first preparing site 4 as $|P\rangle$, and then applying two controlled-phase operations between qubits 2-4 and 3-4. We use the spatial degree of freedom $|u\rangle$ and $|d\rangle$ of photons to carry the qubits 1' and 3'. The corresponding tensor matrices are respectively $B[|u\rangle_{1'}] = B[|P'\rangle_{3'}] = \hat{H}$ and $B[|d\rangle_{1'}] = B[|M'\rangle_{3'}] = \hat{H}\hat{Z}$, where we denote $|P'\rangle = \frac{1}{\sqrt{2}}(|u\rangle + |d\rangle)$ and $|M'\rangle = \frac{1}{\sqrt{2}}(|u\rangle - |d\rangle)$. The four-photon six-qubit state is thus as follows

$$|\psi_6\rangle = |H\rangle_4 |\mu\rangle_{121'} |\nu\rangle_{33'} + |V\rangle_4 \hat{Z}_2 |\mu\rangle_{121'} \hat{Z}_3 |\nu\rangle_{33'}, \quad (4)$$

where

$$|\mu\rangle = c^2|HHu\rangle + cs|VHu\rangle + cs|VHd\rangle - s^2|VVd\rangle; \quad |\nu\rangle = \frac{1}{2}(c|Hu\rangle + s|Vd\rangle) \quad (5)$$

and $\hat{Z}_i$ represents the Pauli matrix $\sigma_z$ applied on qubit $i$.

## Experimental preparation of the resource states

In our experiment, entangled photons are created by using type-II parametric down conversion[31]. Femto-second laser pulses ($\approx 200$ fs, 76 MHz, 788nm) are converted to ultraviolet pulses through a frequency doubler $LiB_3O_5$ crystal and the ultraviolet laser pulse passes through two nonlinear crystals (BBO), generating two pairs of photons in path 1-2 and 3-4. The observed two-fold coincident count rate is about $5.4 \times 10^4 / s$.

First, let's consider the preparation of the four-qubit cluster state (see Fig. 2a). By placing half-wave plates, we prepare the initial two-photon states as $\frac{1}{\sqrt{2}}(|H\rangle_1|P\rangle_2 + |V\rangle_1|M\rangle_2)$ and $\frac{1}{\sqrt{2}}(|H\rangle_3|P\rangle_4 + |V\rangle_3|M\rangle_4)$. Then, two polarization dependent beam splitter cubes ($PBC_a$) with $T_v = (s/c)^2$ and $T_h = 1$ are placed in paths 2 and 3, where $T_v$ ($T_h$) represents the transmission probability for the vertical (horizontal) polarization. This will transform the entangled two-qubit states into

$$\frac{1}{\sqrt{2}}[|H\rangle_1 (c|H\rangle + s|V\rangle)_2 + |V\rangle_1 (c|H\rangle - s|V\rangle)_2] \quad (6)$$

and $\frac{1}{\sqrt{2}}(c|H\rangle_3|P\rangle_4 + s|V\rangle_3|M\rangle_4)$. According to the method in Ref. 32-37, we can apply a C-phase gate between qubit 1 and 4 through a combination of $PBCs$, which includes an overlapping $PBC_a$ ($T_v=1/3$ and $T_h=1$), and a $PBC_b$ ($T_v = 1$ and $T_h = 1/3$) in each path of 1 and 4. Note that the function of $PBC_b$ in path 1 is to adjust the amplitude of $|H\rangle_1$ and $|V\rangle_1$. By removing the $PBC_b$ in path, we can prepare exactly the state $|\psi_4\rangle$.

Next, let's consider the preparation of the six-qubit state $|\psi_6\rangle$ (see Fig. 2b). Based on the created state $|\psi_4\rangle$, we exchange the labels of the outputs 1 and 2. Then, we let each of the photons 1 and 3 enter a polarization beam splitter ($PBS$). Since a $PBS$ transmits $H$ and reflects $V$ polarization, $H$-polarized photons will follow one path and $V$-polarized photons will follow the other. After post-selection, the final state will be converted to exactly $|\psi_6\rangle$ if we denote the first path as $|u\rangle_{i'}$ and the latter path as $|d\rangle_{i'}$, where $i=1,3$. In the experiment, the spatial qubits play the role of reading out the results, and we need to measure the spatial qubits in the $X$, $Y$ bases, which requires matching different spatial modes on a common beam splitter. We have designed a special crystal combining a $PBS$ and a beam splitter, and then we have used Sagnac-ring interferometers[38, 39] to construct the required single-photon interferometers, which can be stable for about 10 hours.

For the four-qubit state, we have extracted its density matrix by the method of over-complete state tomography and obtained $F = \langle\psi_4|\rho|\psi_4\rangle = 0.74 \pm 0.01$ from the density matrix of the experimental state; for the six-qubit state, we have also calculated its fidelity $F = \langle\psi_6|\rho|\psi_6\rangle = 0.73 \pm 0.01$ by measuring 36 measurement settings in the local decomposition of $|\psi_6\rangle\langle\psi_6|$ (see Supplementary Information for the details). Furthermore, we verify that the entanglement properties of our resource

states are much different from cluster states by measuring both two-point correlations and local entanglement in the states. From Table I, we can conclude that, although the experimental data is non-ideal, none of the tested two-point correlations equals zero and none of the local entropy equals 1, clearly showing that our states are not cluster states.

## Single-qubit rotations

To realize an arbitrary $SU(2)$ single-qubit rotation, we measure the qubits of the four-qubit state $|\psi_4\rangle$ in the basis $B(\zeta) = \{|\zeta_0\rangle, |\zeta_1\rangle\}$, where $|\zeta_0\rangle = s|H\rangle + i\tan\frac{\zeta}{2}|V\rangle$ and $|\zeta_1\rangle = c|H\rangle - is\cot\frac{\zeta}{2}|V\rangle$. For simplicity, we define the outcome $r_j$ to be 0 if the measurement result is $|\zeta_0\rangle$, and as 1 if the result is $|\zeta_1\rangle$. We first consider the case when all the measurement outcomes are 0. This will reduce the success probability of the gate for each step of measurement, but it suffices as a proof-of-principle to demonstrate the single-qubit gate. In each measurement step, single qubit rotation $R_z(\zeta) = \exp(-i\zeta\sigma_z/2)$, followed by a Hadamard operation, can be implemented. Based on the generated four-qubit state, we perform consecutive measurements $B_1(\alpha)$, $B_2(\beta)$, $B_3(\gamma)$ on the physical qubits 1, 2, 3. By doing so, the input state of the correlation system $|\psi_{in}\rangle_c = |+\rangle$ is transformed into $|\psi_{out}\rangle_c = \hat{H}R_z(\gamma)R_x(\beta)R_y(\alpha)|+\rangle$ where $R_x(\beta) = \exp(-i\beta\sigma_x/2)$ (see Fig. 1a). Following the map induced by the tensor matrix $B$, the output state of the physical state is $|\psi_{out}\rangle_p = R_z(\gamma)R_x(\beta)R_y(\alpha)|+\rangle|_{0\to H, 1\to V}$. In Table. II, we show the experimental fidelities of six states, the average of which is $0.86 \pm 0.01$, clearly above the classical threshold[40] 2/3. More results of the single-qubit gate can be found in the Supplementary Information.

A distinct feature of MQC in correlation space is the strategy to compensate the randomness of measurement outcomes. In cluster state quantum computation, the wrong measurement outcome only induces by-product Pauli operators, which can be compensated by using feed-forward control of future measurement basis. The by-product operators in MQC in correlation space can be different from Pauli matrices, and the simple feed-forward technique does not work in general any more. Fortunately, it has been proved that the introduced errors can still be efficiently corrected in a bounded number of steps with the trial-until-success strategy[10-12].

As a proof-of-principle, we compare the success probability of implementing the rotation $\hat{H}R_z(\alpha)|+\rangle$ with a two-qubit state $|\lambda_{34}\rangle = c|H\rangle_3|P\rangle_4 + s|V\rangle_3|M\rangle_4$ and with the four-qubit state $|\psi_4\rangle$. Based on the two-qubit state, we measure qubit 1 in the basis $B(\alpha)$. When the outcome $r_1 = 0$, we obtain the desired single-qubit rotation $\hat{H}R_z(\alpha)|+\rangle$ with a success probability $p_s(\alpha)$. However, when $r_1 = 1$, we obtain a rotation $\hat{H}R_z(\alpha')|+\rangle$ but with a wrong angle $\tan(\alpha'/2) = -(1/3)*\cot(\alpha/2)$. Based on the four-qubit state, we first measure qubit 1 in the basis $B(\alpha)$. When obtaining $r_1 = 0$, we measure the qubits 2 and 3 in the $Z$ basis, resulting an output state $\hat{X}^{r_3}\hat{Z}^{r_2}\hat{H}R_z(\alpha)|+\rangle$. When $r_1 = 1$, we measure the second qubit in the $Z$ basis with the outcome $r_2$ and the third qubit in the basis $B[(-1)^{r_2}(\alpha-\alpha')]$. When $r_3 = 0$, with a probability $p_s(\alpha-\alpha')$, we obtain the output state $\hat{Z}^{r_2}\hat{H}R_z(\alpha)|+\rangle$. The final result is equivalent to the desired rotation up to a Pauli by-product operator. Thus, with two more qubits, we can boost the success probability of the rotation from $p_s(\alpha)$ to $p_s(\alpha) + [1 - p_s(\alpha)] \cdot p_s(\alpha-\alpha')$ (see Supplementary Information for the case of more qubits).

The experimental success probability in the four-qubit case is achieved by measuring in the appropriate basis and then adding the probability of each successful measurement branch. The theoretical and experimental success probabilities are shown in Fig. 3a, from which we can see that, the experimental data is well consistent with the theoretical curve and the success probability has indeed been increased with compensation.

## Two-qubit entangling gates

Besides single-qubit rotations, a two-qubit entangling gate is required to demonstrate universal quantum computing. Based on the six-qubit state $|\psi_6\rangle$, we have realized a two-qubit controlled phase gate. We first measure qubit 1 in the basis $B(\alpha)$. When $r_1 = 0$ (we only consider this case in the following text, the other outcome just corresponds to a different input), this measurement initializes the input state of the correlation system as $\hat{H}R_z(\alpha)|+\rangle_t \otimes |+\rangle_c$, where $t$ and $c$ denote the target and control logical qubit corresponding to two computational wires (1-2-1' and 3-3'). Then, we measure qubit 2 and 3 in the basis $B(\frac{\pi}{2}) = \{s|H\rangle + ic|V\rangle, c|H\rangle - is|V\rangle\}$. (a) Once having the outcome $r_2 = r_3 = 0$, we measure qubit 4 in the $Y$ basis $\{|H\rangle + i|V\rangle, |H\rangle - i|V\rangle\}$ and transform the logical state carried by the correlation systems from $|\psi_{in}\rangle = \hat{H}R_z(\alpha)|+\rangle_t \otimes |+\rangle_c$ to $|\psi_{out}\rangle = (\hat{H} \otimes \hat{H}) \cdot (\hat{Z} \otimes \hat{Z})^{r_4} \cdot CZ|\psi_{in}\rangle$, where $CZ = I \otimes |0\rangle\langle 0| + \hat{Z} \otimes |1\rangle\langle 1|$ is the controlled phase gate. The tensor matrices corresponding to qubits 1' and 3' map the logical output state to the physical output state[28] carried by qubits 1' and 3' as

$$(I \otimes \hat{H}) \cdot (\hat{Z}_{1'} \otimes \hat{Z}_{3'})^{r_4} \cdot (|0\rangle|+\rangle - i\tan\frac{\alpha}{2}|1\rangle|-\rangle)_{1'3'} |_{0 \to u, 1 \to d}. \qquad (8)$$

(b) If $r_2 \neq 0$ or $r_3 \neq 0$, we measure qubit 4 in the $Z$ basis and decouple two 1D wires. After the measurement, a local rotation error will appear on each logical qubit. The same as in the single-qubit gate, we can efficiently correct them by applying the trial-until-success strategy if we have more qubits.

In the experiment, when $r_2 = r_3 = 0$, we characterize the output state of the two-qubit gate by the method of state tomography. We collect the experimental data for 600s for each of 36 combinations of the measurement basis $\{|H'\rangle, |V'\rangle, |P'\rangle, |M'\rangle, |R'\rangle, |L'\rangle\}$, where $|R'\rangle = \frac{1}{\sqrt{2}}(|u\rangle + i|d\rangle)$ and $|L'\rangle = \frac{1}{\sqrt{2}}(|u\rangle - i|d\rangle)$, and then estimate the density matrix with the maximum likelihood technique. The theoretical and experimental density matrices of physical output states when $\alpha = 0$, $r_4 = 0$ and $\alpha = \pi/3$, $r_4 = 0$ are shown in Fig. 3. The states are in good agreement with the ideal states, which can be seen from their fidelities: $0.88 \pm 0.02$ and $0.84 \pm 0.03$. More data when $r_2 = r_3 = 0$ and the detailed fidelities when $r_2 \neq 0$ or $r_3 \neq 0$ are shown in the Supplementary Information.

## Deutsch's algorithm

Deutsch's algorithm represents an interesting instance of demonstrating the power of quantum computation over classical computation[41, 42]. Here we experimentally demonstrate the implementation of Deutsch's algorithm based on the six-qubit state $|\psi_6\rangle$. Deutsch's algorithm, also known as the Deutsch-Jozsa algorithm[43], allows one to distinguish two different types of function $f(x)$ implemented by an oracle in a black box. The function $f(x)$ with an $N$-bit binary input $x$ is constant if it returns the same value (either 0 or 1) for all possible inputs and balanced if it returns 0 for half of the inputs and 1 for the other half. With classical methods, in some cases we have to query this oracle as many as $2^{N-1}+1$ times. However, only one query is required in all cases by using quantum computation[43]. In the two-qubit version, the applied algorithm can be demonstrated as $|x\rangle|y\rangle \rightarrow |x\rangle|y \oplus f(x)\rangle$, where $|x\rangle$ is the query input qubit and $|y\rangle$ is the ancilla input qubit. Preparing $|x\rangle|y\rangle$ as $|+\rangle|-\rangle$, we can get the output states $[(-1)^{f(0)}|0\rangle+(-1)^{f(1)}|1\rangle]|-\rangle$. Finally, we apply two Hadamard operations on the two output qubits. If $f(x)$ is constant, the final result will be $|H\rangle|V\rangle$, while if it is balanced, the final result will be $|V\rangle|V\rangle$. Therefore, now we can determine the types of $f(x)$ from the output states. There are two types of constant function, namely $f_1 = I \otimes I$ and $f_2 = I \otimes \sigma_x$. Also, there are two types of balanced function, namely $f_3 = g_{cnot}$ and $f_4 = g_{cnot}(I \otimes \sigma_x)$, where $g_{cnot}$ represents a controlled-not gate. Since $f_2$ and $f_4$ can be achieved from $f_1$ and $f_3$ by single-qubit local operations, in the following we consider only the functions $f_1$ and $f_3$.

In the experiment, we use the wire 3-3' of $|\psi_6\rangle$ to carry the query logical qubit and 1-2-1' of $|\psi_6\rangle$ for the ancilla logical qubit. In the following, we consider the implementation of Deutsch's algorithm when $r_1 = r_2 = r_3 = r_4 = 0$. The measurement of qubit 1 in the basis $B(\pi)$ and qubits 2, 3, 4 in the basis $B(0)$ performs the transformation $H_q \otimes R_z(\pi)_a$. Moreover, considering the map of logical qubits from correlation systems into the physical qubits the final transformation of the inputs will be $(\hat{H}_q \otimes \hat{H}_a)(I_q \otimes R_z(\pi)_a)$, which implements the preparation of ancilla qubit $|-\rangle$, the constant function and the readout of the qubits (see Fig. 4a). Similarly, by measuring qubit 1 in the basis $B(\pi)$ and qubits 2, 3, 4 in the basis $B(\pi/2)$, we can implement the gate $(\hat{H}_q \otimes \hat{H}_a)g_{cnot}(I_q \otimes R_z(\pi)_a)$ on the inputs, which is the case

when the function is balanced (see Fig. 4b). Here the measurement of qubits 2, 3 and 4, together with the detector signal corresponding to $r_1 = r_2 = r_3 = r_4 = 0$ should be viewed as the entire black box. In the experiment, the success probability of recognizing the type of $f(x)$ is as large as $99\% \pm 1\%$ for $f_1$ and $75\% \pm 2\%$ for $f_3$. The non-ideal probability is mainly caused by the non-perfect resource state and single-photon interferometers in the setup. In the Supplementary Information, we have discussed in detail the cases when some measurement outcomes are not zero.

## Discussion

There are many open questions, both theoretical and experimental, worth investigating in the future. In MQC with cluster states, efficient linear optical quantum computation is possible with detector efficiency above 1/2 and arbitrary small source efficiency[44, 45]. With novel resource states for quantum computation in correlation space, computation will probably require more qubits. Nevertheless, it has been proved that bonds between two particles can be easier to create if the required entanglement is below certain levels[15-17], which is related to the resource requirements[44-46]. It would be interesting to investigate what conditions are required to make the latter scheme scalable. Moreover, in our experiment we have used the spatial modes of photons as read-out qubits in the two-qubit entangling gate. It is useful to grow a larger resource state this way; however, in the current setup, the spatial qubit is limited to connect the polarization qubit of the same photon only. More scalable ways of hyper-entanglement should be investigated. Third, the future implementations of photonic quantum computer will likely require integrated waveguides[18-20]. The pioneering works of integrated waveguides including ultra-stable interferometer[18, 19] and precise optical phase control[20] have been demonstrated. The combination of the present scheme with integrated waveguides will further advance the development of optics quantum computing. Finally, the feed-forward rule also warrants further study, and it would be desirable to combine the feed-forward technique[8, 47] with the proof-of-principle demonstration.

**Acknowledgements** We thank H. J. Briegel for valuable suggestions, and J. Eisert for helpful discussions. This work is supported by the NNSF of China, the CAS, the National Fundamental Research Program (under Grant No. 2006CB921900), the Fundamental Research Funds for the Central Universities. The research at Innsbruck is supported by the FWF (J.-M. C. through the Lise Meitner Program, SFB-FoQuS).

Correspondence and requests for materials should be addressed to Z.-B. C (zbchen@ustc.edu.cn) or J.-W. P. (pan@ustc.edu.cn).

**Author contributions** W.-B.G., J.-M. C., Z.-B. C. and J.-W.P. conceived the research; W.-B.G., X.-C.Y., H.L., P.X., T. Y., C.-Y. L. and Y.-A. C. carried out the experiment; W.-B.G., X.-C.Y., J.-M. C., H.L. and P.X. analysed the data; W.-B.G., J.-M. C., C.-Y. L. Y.-A. C. and J.-W.P. wrote the paper; J.-W.P. and Z.-B.C. supervised the whole project.

**Table I**: **Comparison of the prepared states, the ideal states and the corresponding cluster states.** Here the two-point correlation between qubit $i$ and $j$ is defined as $Q_{ab}^{ij}(|\Psi\rangle) = \langle\Psi|\hat{a}_i \otimes \hat{b}_j|\Psi\rangle - \langle\Psi|\hat{a}_i|\Psi\rangle\langle\Psi|\hat{b}_j|\Psi\rangle$, where $\hat{a}_i$ and $\hat{b}_j$ are respectively Pauli matrices acting on qubits $i$ and $j$. Local entanglement between one single qubit $i$ and the other qubits is quantified by the local entropy $E(\rho_i) = 2(1 - Tr\rho_i^2)$, where $\rho_i$ is the reduced density matrix of qubit $i$. In spite of the imperfect results, it is clear to see the difference of the prepared states from cluster states.

| State types | Four-qubit states | | | | Six-qubit states | | | | | |
|---|---|---|---|---|---|---|---|---|---|---|
| | $Q_{XX}^{13}$ | $E(\rho_1)$ | $E(\rho_2)$ | $E(\rho_3)$ | $Q_{XZ}^{24}$ | $Q_{ZX}^{34}$ | $E(\rho_1)$ | $E(\rho_2)$ | $E(\rho_3)$ | $E(\rho_4)$ |
| Cluster states | 0 | 1 | 1 | 1 | 0 | 0 | 1 | 1 | 1 | 1 |
| Ideal states | 0.375 | 0.75 | 0.5625 | 0.75 | 0.433 | 0.375 | 0.75 | 0.75 | 0.75 | 0.9375 |
| Prepared states | 0.26 (0.01) | 0.93 (0.01) | 0.63 (0.01) | 0.88 (0.01) | 0.41 (0.02) | 0.32 (0.02) | 0.62 (0.02) | 0.80 (0.02) | 0.67 (0.02) | 0.90 (0.02) |

**Table II**: **The fidelities of the output states of the single-qubit rotation.** The first three qubits of the four-qubit cluster state are measured in basis $B_1(\alpha)$, $B_2(\beta)$, $B_3(\gamma)$. $|\psi_{out}\rangle_c$ and $|\psi_{out}\rangle_p$ represent the output states in the correlation space and the physical world, respectively. Here $|I_0\rangle = \frac{1}{\sqrt{2}}(|0\rangle + i|1\rangle)$ and $|I_1\rangle = \frac{1}{\sqrt{2}}(|0\rangle - i|1\rangle)$.

| $\alpha$ | $\beta$ | $\gamma$ | $|\psi_{out}\rangle_c$ | $|\psi_{out}\rangle_p$ | fidelity |
|---|---|---|---|---|---|
| 0 | 0 | 0 | $|0\rangle$ | $|P\rangle$ | $0.92\pm0.01$ |
| $\pi$ | $\pi$ | 0 | $|1\rangle$ | $|M\rangle$ | $0.72\pm0.02$ |
| $\pi/2$ | $\pi/2$ | $\pi/2$ | $|+\rangle$ | $|H\rangle$ | $0.80\pm0.02$ |
| $-\pi/2$ | $\pi/2$ | $\pi/2$ | $|-\rangle$ | $|V\rangle$ | $0.91\pm0.01$ |
| $\pi$ | $\pi$ | $\pi/2$ | $|I_0\rangle$ | $|L\rangle$ | $0.93\pm0.01$ |
| $\pi$ | $\pi$ | $-\pi/2$ | $|I_1\rangle$ | $|R\rangle$ | $0.86\pm0.02$ |

**Figure Captions:**

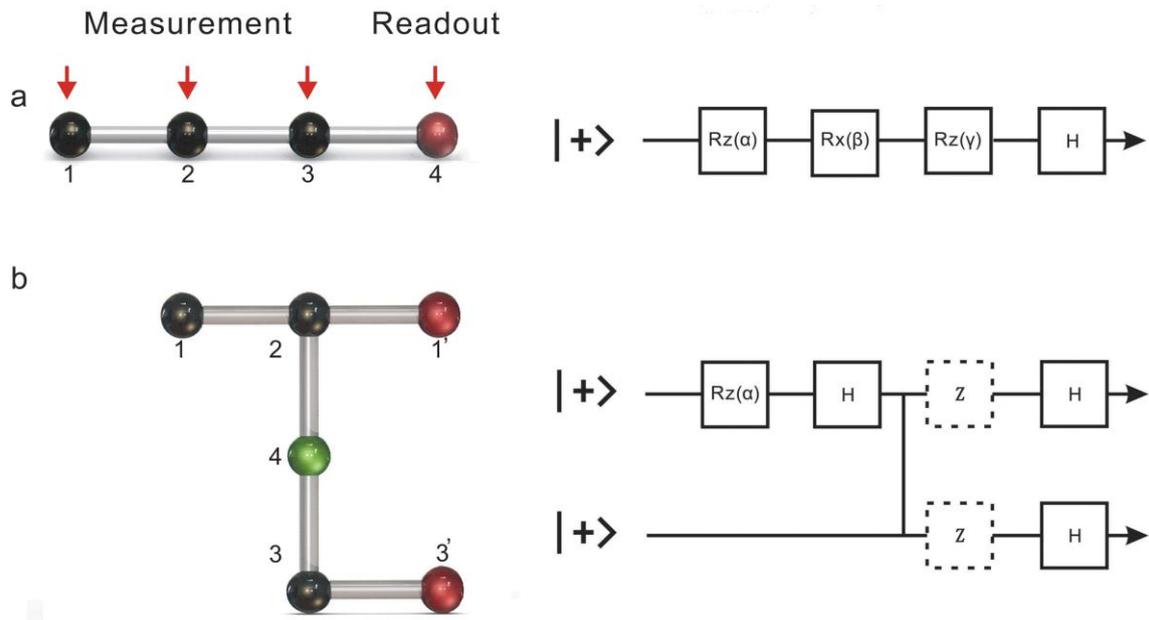

**Figure 1: Entangled states and their corresponding circuits. a.** A four-qubit state for the implementation of single qubit rotations. The black dots are associated with the tensor matrices $A[|H\rangle] = \hat{H}\cos\theta$ and $A[|V\rangle] = \hat{H}\hat{Z}\sin\theta$, while the red dots are associated with $B[|H\rangle] = \hat{H}$, $B[|V\rangle] = \hat{H}\hat{Z}$. The measurements on the first three qubits will implement the quantum circuit on the right. **b.** A six-qubit state for the implementation of a two-qubit entangling gate. The horizontal lines (1-2-1' and 3-3' ) represent two computational wires corresponding to two logical qubits. To couple them, we prepare site 4 as $\frac{1}{\sqrt{2}}(|H\rangle + |V\rangle)$ and then apply two controlled-Z operations on qubits 2-4 and 3-4. The measurements on the qubits 1-2-3-4 will implement a C-Phase gate shown on the right. The $\hat{Z} \otimes \hat{Z}$ operation will depend on the measurement result of qubit 4, and thus they are drawn in dashed squares.

a

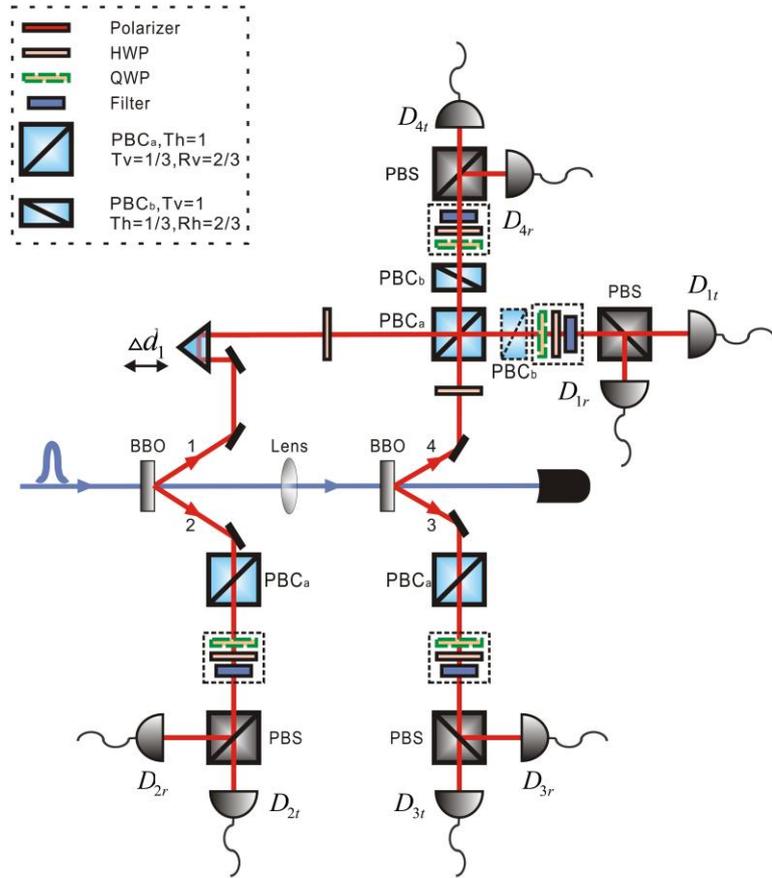

b

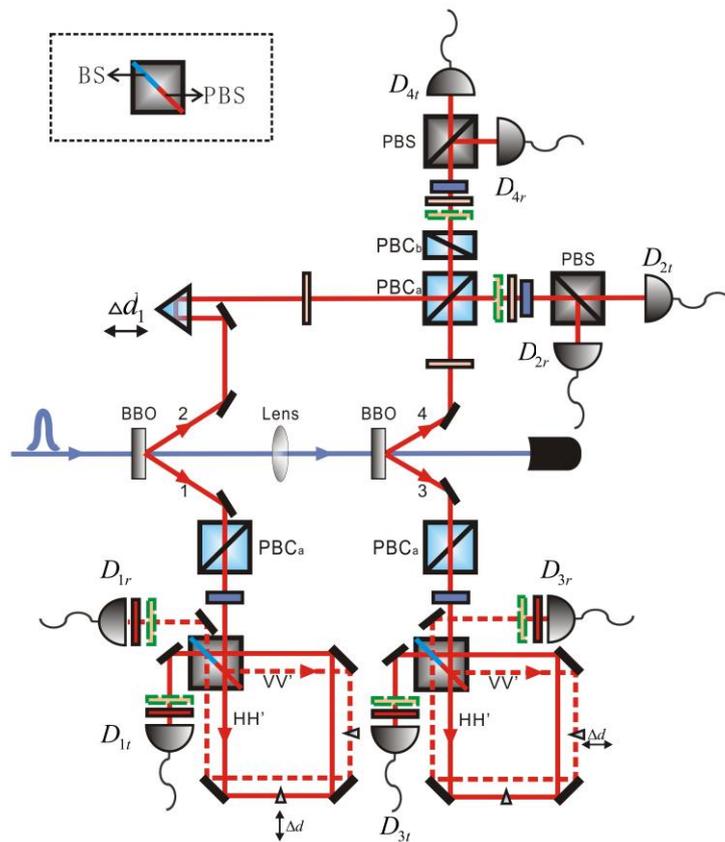

**Figure 2: Experimental setup for the generation of the four-qubit state $|\psi_4\rangle$ and the six-qubit state $|\psi_6\rangle$. a.** An ultra-violet pulse passes through two BBO crystals to create two pairs of entangled photons. Then, half-wave plates (HWP) and a series of polarization dependent beam splitter cubes (PBCs) have been used to create the desired state. Prism $\Delta d1$ is used to ensure that the input photons arrive at the PBCs at the same time. Furthermore, every output is spectrally filtered ($\Delta\lambda_{FWHM} = 3.2nm$) to ensure good temporal overlap. A combination of HWP, quarter-wave plates (QWP) and polarization beam splitter (PBS) has been used to implement the measurement setups. **b.** By exchanging the labels of 1 and 2 of $|\psi_4\rangle$, and letting the photons 1 and 3 enter two PBSs, the desired six-qubit state $|\psi_6\rangle$ can be prepared. The ultra-stable Sagnac-ring interferometers have been used to measure the spatial qubits. In the interferometer, the specially designed beam-splitter cubes are half PBS-coated and half beam splitter coated and high-precision small-angle prisms are used to satisfy fine adjustments of the relatively delay of the two different paths.

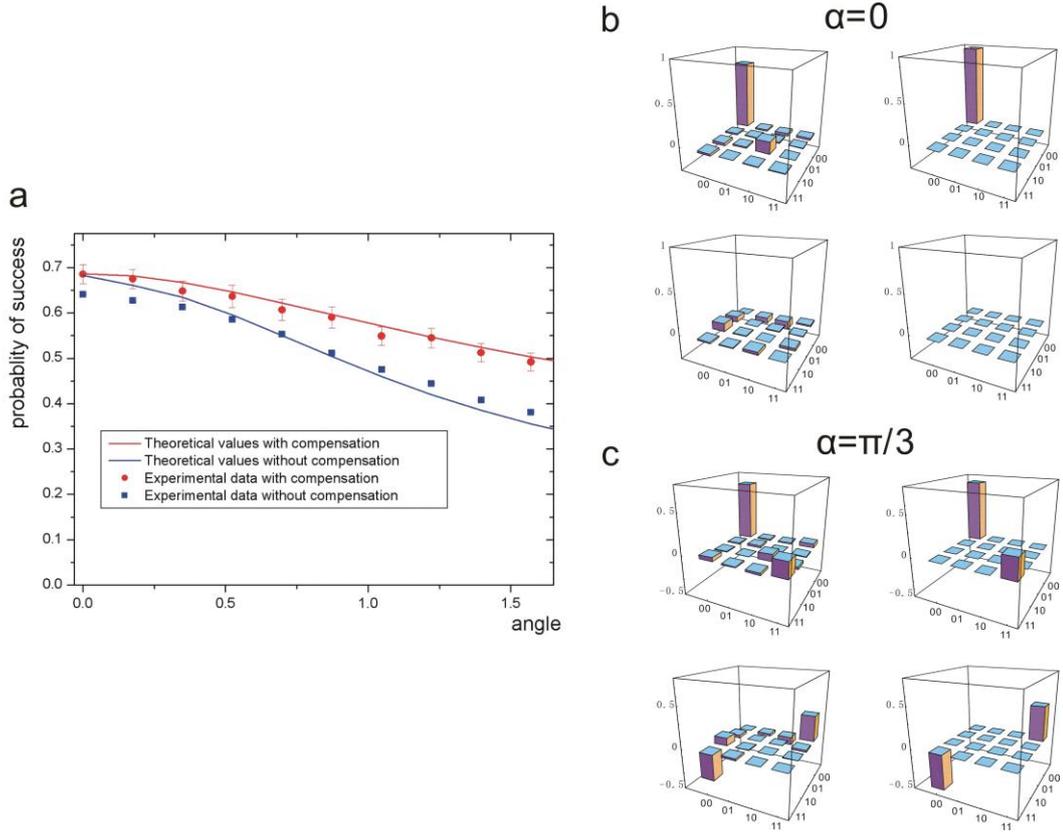

**Figure 3: The success probability single-qubit rotation and the density matrices of the output states of the entangling two-qubit gate. a.** The theoretical and experimental success probability of single-qubit rotation with and without compensation of the rotation error. The blue (red) line represents the theoretical success probability under the white noise model in the two-qubit (four-qubit) case. Here the two states are written as $\rho_2 = p|\lambda_{34}\rangle\langle\lambda_{34}| + (1-p)I_2/4$ and $\rho_4 = p'|\psi_4\rangle\langle\psi_4| + (1-p')I_4/16$, where $I_2(I_4)$ denotes two-qubit (four-qubit) identity matrix; $p = (4f_2 - 1)/3$; $p' = (16f_4 - 1)/15$ and $f_2(f_4)$ is the fidelity of the two-qubit (four-qubit) state, that is, 0.90 (0.73). The blue (red) dots represent the experimental values in the two-qubit (four-qubit) case, which are achieved by measuring the states for 5s (300s). From the figure, we can see that the success probability has been enhanced by using two more qubits. **b. c.** The density matrices of the output states of the entangling two-qubit gate. The left column shows the real and the

imaginary part of the experimental output state density matrices, while the right column shows the theoretical density matrices. In cases b and c, qubit 1 is respectively measured in bases $B(0)$ and $B(\pi/3)$. All the matrices are achieved in the case when the outcome of the measurements is $r_1 = r_2 = r_3 = r_4 = 0$.

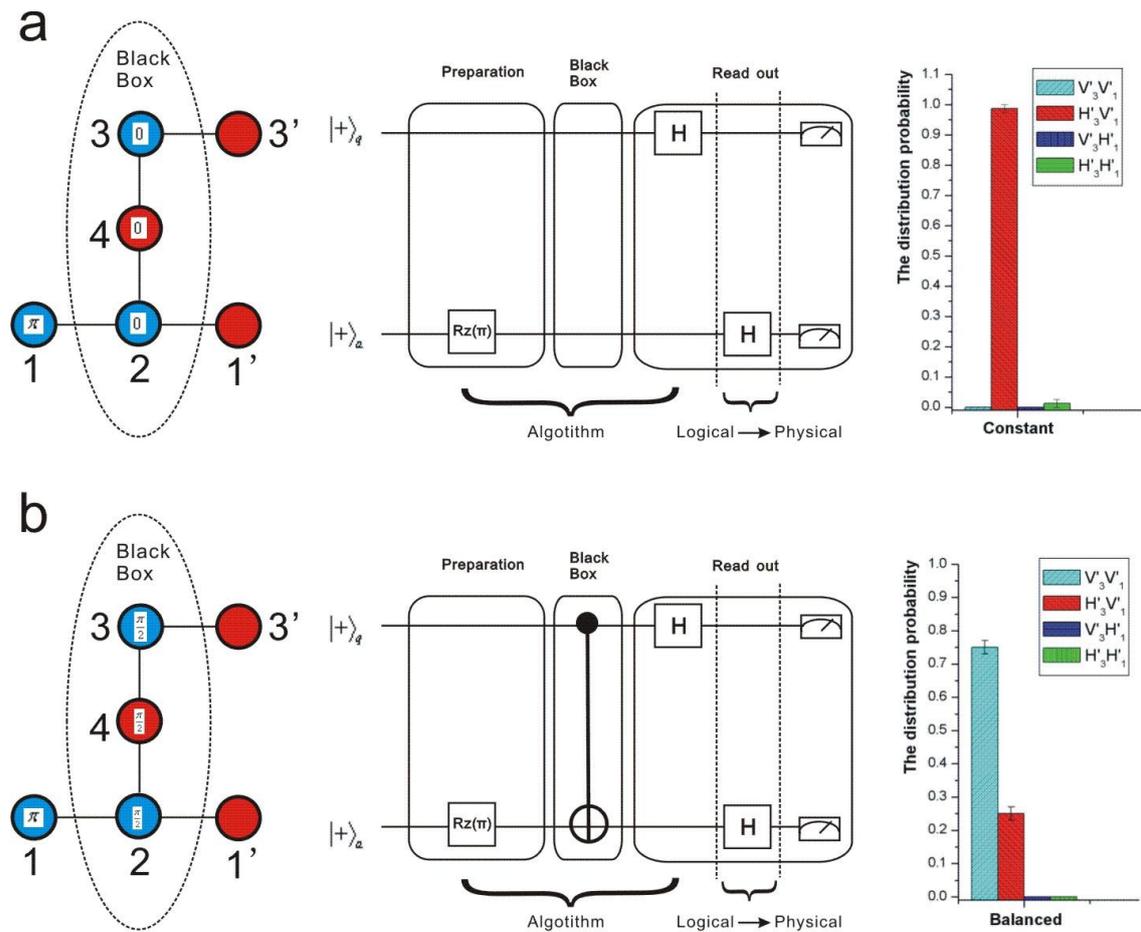

**Figure 4: The theoretical design and the experimental results of implementing Deutsch's algorithm.** The wire 3-3' represents the query logical qubit, and the wire 1-2-1' represents the ancilla logical qubit. In **a** and **b**, the measurement patterns on the physical qubits (left) implement the algorithm on the logical qubits (right). The angles $\alpha$ represent the measurement basis $B(\alpha)$ of each physical qubit. In the circuit, the operation $(I \otimes \hat{H})$ represents

the mapping of the qubits from the correlation space to the realistic physical world. The final results are carried by the qubits 3' and 1'. The success probability of recognizing the constant function is $99\%\pm1\%$, while for the balanced function, this probability is $75\%\pm2\%$.